\documentclass[graybox]{svmult}

\usepackage{type1cm}        
\usepackage{makeidx}         
\usepackage{graphicx}        
\usepackage{multicol}        
\usepackage[bottom]{footmisc}

\usepackage{newtxtext}       %
\usepackage[varvw]{newtxmath}       

\makeindex             

\begin{document}

\title*{Time-continuous microscopic pedestrian models: an overview}
\author{Raphael Korbmacher, Alexandre Nicolas, Antoine Tordeux and Claudia Totzeck}
\institute{Raphael Korbmacher \at School of Mechanical Engineering and Safety Engineering, University of Wuppertal, Gau{\ss}str.~20, Germany, \email{korbmacher@uni-wuppertal.de}
\and Alexandre Nicolas \at Institut Lumi\`ere Mati\`ere, CNRS \& Universit\'e Claude Bernard Lyon 1, 69622, Villeurbanne, France \email{alexandre.nicolas@polytechnique.fr}
\and Antoine Tordeux \at School of Mechanical Engineering and Safety Engineering, University of Wuppertal, Gau{\ss}str.~20, Germany, \email{tordeux@uni-wuppertal.de}
\and Claudia Totzeck \at IMACM, University of Wuppertal, Gau{\ss}str.~20, Germany, \email{totzeck@uni-wuppertal.de}
}
%
%
\maketitle

\abstract*{We provide an overview of time-continuous pedestrian models, with a focus on data-driven modelling. 
The presentation of the various models puts emphasis on their mathematical aspects and their different components. 
Starting from pioneer, reactive force-based models, we move forward to modern, active pedestrian models with sophisticated collision-avoidance techniques and anticipation processes, which can ultimately be handled as optimisation problems. 
We then discuss methods used for data-based calibration of model parameters, hybrid approaches incorporating neural networks, and purely data-based models fitted by deep learning. To conclude, we outline
some development perspectives of modelling paradigms that we expect to grow in the coming years.}

\abstract{We give an overview of time-continuous pedestrian models with a focus on data-driven modelling. Starting from pioneer, reactive force-based models we move forward to modern, active pedestrian models with sophisticated collision-avoidance and anticipation techniques through optimisation problems. The overview focuses on the mathematical aspects of the models and their different components. We include methods used for data-based calibration of model parameters, hybrid approaches incorporating neural networks, and purely data-based models fitted by deep learning. The conclusion outlines some development perspectives that we expect to grow in the coming years.}


\section{Introduction}\label{sec:introduction}

Video recordings of people moving around in a train station at rush hour show some striking similarities with streams of liquid. This is reflected in the fact that the origin of mathematical pedestrian models stems from partial differential equations modelling the flow of gas or fluids \cite{helbing1992fluid,henderson1971statistics} with Navier-Stokes or Boltzmann-type dynamics. The formulation in terms of partial differential equations is useful for mathematical analysis, numerical simulation, and optimization and the behaviour of large crowds is well-approximated. However, as these models consider averaged quantities, their information is only statistical. Indeed, the particular position or velocity of each pedestrian is not known. The latter is needed if the objective is to understand interaction of small groups or just two pedestrians for example to model collision avoidance. To fill this gap, microscopic models were proposed. 

The seminal work by Hirai and Tarui \cite{hirai1975simulation} stays close to the physical laws underlying the macroscopic descriptions but introduces forces to model pedestrian behaviors. This class of models was democratised at the turn of the century by Helbing and M\'olnar \cite{helbing1995social} under the concept of \textit{social forces} representing human reactions, psychological and mental processes, and other stimuli. Moreover, an acceleration term prescribing the relaxation towards the desired velocity of each pedestrian is incorporated. The forces acting on one pedestrian are the sum of the forces resulting from the interactions of this pedestrian with peers or unknown pedestrians, as well as interactions with walls. This allows to reduce the modelling to forces of binary interactions between pedestrians or wall points and is consistent with the macroscopic models. In fact, rigorous relations can be shown, for example, in terms of the mean-field scaling \cite{albi2019vehicular}, and thus analytical results such as stationary states or time-invariant behaviours proved in the mean-field limit can be transferred to the microscopic dynamics. Furthermore, the simple interaction rules allow for efficient implementation and show realistic behaviours such as formation of lanes or gridlocks. Moreover, they lay the ground for optimization approaches. 

The huge amount of data available nowadays in combination with optimization permits alternative approaches to pedestrian modelling and simulation. On the one hand, the formulation as an inverse problem allows for parameter calibration of the models at hand. On the other hand, data-driven modelling, where forces or even the whole trajectories of pedestrians are predicted with the help of neural network architectures, are new streams in pedestrian research. In fact, we see great potential in the formulation of hybrid models that are feasible for mathematical investigation while curing artefacts from physical models that lead to unrealistic behaviour in social interactions. 

In the following, we give an overview of microscopic pedestrian models beginning with a summary of classical force-based models. Then we discuss refinements such as velocity and anticipation-based models before we highlight recent ideas on data-based calibration and prediction with the help of neural networks. 
General overviews of pedestrian models including macroscopic approaches and discrete algorithms can be found in the surveys \cite{aylaj2020unified,bellomo2022towards,chraibi2018modelling,cristiani2014multiscale,martinez2017modeling,van2021algorithms}.
A deep introduction to the microscopic modelling of pedestrian crowds is proposed in the book by Maury and Faure \cite{maury2018crowds}. 
A survey comparing physics-based models and data-based algorithms for pedestrian trajectory prediction is given in \cite{korbmacher2022review}, while \cite{vermuyten2016review} provides a literature review of optimisation models for pedestrian evacuation and design problems.


\section{Force-based models}\label{sec:force}

Force-based models assume that pedestrian motion is governed by ordinary differential equations of second order with a superposition of external forces \cite{chraibi2011force}. 
Each pedestrian is represented by a particle whose kinematic is continuous in time and space. 
In the simplest case, the pedestrians are represented by point masses.  
More sophisticated models use volume exclusion with circular shape or even  ellipsoids whose size can be velocity-dependent \cite{chraibi2010generalized,song2018simulation}. 

Force-based models are borrowed from the physics of systems of interacting particles. However, instead of defining interactions resulting from shocks or fields of attraction and repulsion obeying Newton's law, the interactions of pedestrian models result from \textit{social forces}. These forces aim to avoid collisions and, based on concepts of proxemics, to keep a certain distance from neighbours.  
Force-based models, and more particularly the social force model of Helbing and Moln\'ar \cite{helbing1995social}, are nowadays the standard simulation tool for engineers.
In the following, we examine the main modelling components of the pioneer force-based model developed by Hirai and Tarui. Afterwards, we provide an overview of modern distance-based and velocity-and-distance-based models.

\subsection{Pioneer force-based model by Hirai and Tarui}
Pioneer works on pedestrian force-based models date back to the contribution of Hirai and Tarui on the simulation of crowd behavior in panic \cite{hirai1975simulation}. 
Considering $N\in \mathbb N$ pedestrians with position $x_i$ and velocity $v_i$, $i=1,\ldots,N$, the model of Hirai and Tarui (HT model) reads
\begin{subequations}\label{eq:HT}
\begin{align}
\frac{d}{dt} x_i &= v_i, \qquad i=1,\dots,N, && x_i(0) = x_{0,i}, \\
m_i\frac{d}{dt} v_i &= -\varphi_i v_i+F_{1i}(t,x,v) + F_{2i}(t,x) + \xi_{i}(t), && v_i(0) = v_{0,i}.
\end{align}
\end{subequations}
Here, $m_i$ and $\varphi_i$ are the pedestrian mass and a coefficient of viscosity, respectively. 
Note that the mass and the viscosity coefficient can be specific to each pedestrian, introducing a kind of heterogeneity in the pedestrian behavior. More details regarding the question of anatomical and behavioural heterogeneity will be discussed further in Sec.~\ref{sub:articulation} and \ref{sub:diff_games}. 
The term $\xi_{i}$ is a random force while $F_{1i}$ and $F_{2i}$ are external forces acting upon the individual $i$ for $i\in\{1,\dots,N\}$. 
The term $F_{1i}$ consists of three components:
\begin{equation}
F_{1i}(t,x,v)=F_{ai}(v_i)+F_{bi}(t,x,v)+F_{ci}(t,x,v)
\end{equation}
where
\begin{itemize}
    \item $F_{ai}$ is an inertial force in the current direction.
    \item $F_{bi}$ is \textit{social} force exerted by the neighbors. The force results from a superposition of distance-based pairwise interaction potentials. The force is attractive for two individuals from a certain distance, while it is repulsive as the two individuals are close to each other. 
    This force can be written as
    \[F_{bi}(t,x,v)=\sum_{j=1}^N\nabla U(x_j-x_i)\,\omega_b(\phi_{ij}),\]
    where $U$ is a repulsive or attractive potential with respect to its argument while $\omega_b$ is an anisotropic factor depending on the bearing angle of the pedestrian $i$ with the pedestrian $j$. The anisotropy gives more weight to pedestrians in the direction of motion.
    \item $F_{ci}$ is a \textit{swarming} force relaxing the pedestrian velocity to the velocities of the neighbors.
    This force reads
    \[F_{ci}(t,x,v)=\sum_{j=1}^N h(|x_j-x_i|)\,\omega_c(\phi_{ij})\,(v_j-v_i).\]
    Here $h$ is a function being constant upon a fixed interaction threshold and zero elsewhere. 
    The swarming force is anisotropic due to $\omega_c$, hence pedestrians in front have more influence on the dynamics of pedestrian $i$.
\end{itemize}
The term $F_{2i}$ represents a force exerted by the environment on the $i$-th pedestrian, including her/his destination and corresponding desired direction. The force consists of four components:
\begin{equation}
F_{2i}(t,x)=F_{wi}(t,x_i)+F_{ki}(t,x_i)+F_{gi}+F_{hi}
\end{equation}
with
\begin{itemize}
    \item $F_{wi}$ is a distance-based force modelling the interaction with walls and obstacles in the environment. This force is systematically repulsive and perpendicular to the walls and obstacles.
    \item $F_{ki}$ models attraction to the destination. This force reads
    \[F_{ki}(t,x_i)=\alpha_i\frac{P-x_i}{|P-x_i|}\]
    where $\alpha_i>0$ is the attraction magnitude and $P$ the location of the destination. 
    \item $F_{gi}$ provides the motion direction close to the exit while $F_{hi}$ is a constant force causing the individual $i$ to move away from a site in case of panic.
    \end{itemize} 
The HT model is a pioneering work in the field of pedestrian modelling. 
However, it includes a rich set of concepts and interaction mechanisms. 
In the following section, we will observe that many of these concepts have been incorporated into modern force-based models.

\subsection{Modern force-based models}

Force-based models mainly rely on interaction terms with the neighboring pedestrians similar to the forces $F_b$ and $F_c$ of the Hirai and Tarui model presented above. In addition, a drift term provides desired directions to the pedestrians. 
A general force-based model is given by
\begin{subequations}\label{eq:FBgeneral}
\begin{align}
\frac{d}{dt} x_i &= v_i, \qquad i=1,\dots,N, && x_i(0) = x_{0,i}, \\
\frac{d}{dt} v_i &= D_i(t,x_i,v_i) + I_i(t,x,v) + B(t,x_i,v_i), && v_i(0) = v_{0,i}.
\end{align}
\end{subequations}
Modern force-based modelling approaches refer to a pseudo-Newtonian framework. Here, in contrast to the HT model, the mass of the agent is no longer taken into account. 
The force $B$ describes the interaction with the environment, i.e., walls or obstacles. 
It is not considered in the following.
The drift term  $D_i$ generally models relaxation of the velocity $v_i$ towards a desired velocity $u_i$ \cite{chraibi2011force,helbing1995social}
\begin{equation}
D_i(t,x_i,v_i)=\tau(u_i(x_i)-v_i).
\end{equation}
Similarly to the force $F_k$ in the HT model, the desired velocity $u_i$ can depend on the current position. 
Yet, a relaxation mechanism acts upon the velocity while the attraction magnitude is constant in the HT model.
The main modelling component of force-based models is the interaction term $I_i$ specifying  social forces which are inspired from proxemic concepts, and kinematic and collision avoidance behaviors. 
The interaction term is a superposition of pairwise forces with the neighbors. These binary interaction forces are usually assumed to be the gradient of a repulsive potential and multiplied by an anisotropic factor modelling for example a vision cone effect with forward interaction range or rational behavior \cite{bailo2018pedestrian}.
The interaction force of distance-based models may be written as
\begin{equation}
I_i(t,x,v)=-\sum_{j=1}^N \nabla U(x_j-x_i) \,\omega(\phi_{ij})
\end{equation}
with $U$ the repulsive potential and $\omega$ the anisotropic factor depending for example on the relative bearing angle with the neighbors $\phi_{ij}$.
In the famous social force model \cite{helbing1995social}, these functions read
\[ U(x)=AB\exp(-|x-\ell|/B)\qquad\text{and}\qquad 
\omega(x)=\left\{\begin{array}{ll}1~~~&\text{if}~|\phi_{ij}|<\kappa\\
c&\text{otherwise}\end{array}\right.\]
where $\kappa$ is the vision cone angle, and $0<1<c$ is a reduced perception factor. 
The parameters $A$ and $B$ are the repulsion rate and distance, respectively. 
Besides, an additional repulsive layer with higher rate and lower repulsion distance accounts for hard-core exclusion in case of contact in the initial model \cite{helbing1995social}.
Although the social force model is simple, with only six parameters ($\tau$, $\kappa$, $c$, $A$, $B$, and $\ell$), it can describe many collective phenomena attributed to pedestrian crowds. 
Prominent examples are jamming and clogging at bottlenecks, stop-and-go in uni-directional flow, lane formation for counter-flow, or strip formation for crossing flow \cite{helbing2005self,helbing2000simulating,helbing1995social}.
In the optimal velocity model \cite{nakayama2005instability}, the potential gradient is an arc-tangent function, and the anisotropic factor simply reads $\omega(x)=1+\cos(x)$. 
This distance-based model is more used for theoretical purpose, i.e., stability analysis \cite{li2010stability,nakayama2008effect}.

The interaction potentials are distance-based in the social force and optimal velocity models.
The velocities of the neighboring pedestrians play no role in the interactions.
This modelling aspect of pedestrian dynamics is questionable. 
More sophisticated models take both, distances and velocities, into account.
In these models, the interaction force is given by
\begin{equation}
I_i(t,x,v)=-\sum_{j=1}^N \nabla_{\!x}\, U(x_j-x_i,v_i,v_j) \,\omega(\phi_{ij}),
\end{equation}
where $U$ is again the interaction potential and $\omega$ an anisotropic factor.
For instance, in the centrifugal and generalised centrifugal force models \cite{chraibi2010generalized,yu2005centrifugal}, the anisotropic factor is $w(x)=\cos(x)$ for $x<\pi/2$ and zero otherwise, while the velocity and distance-based interaction forces read
\[ \nabla_{\!x}\, U_\text{CF}=\frac{\beta(v_j-v_i)^2}{|x_j-x_i|}\qquad\text{and}\qquad 
\nabla_{\!x}\, U_\text{GCF}=\frac{\beta(\eta v_0+v_j-v_i)^2}{|x_j-x_i|+\lambda(|v_i|+|v_j|)},\]
respectively. It is worth observing that the interaction force decays algebraically  with respect to the distance for centrifugal force models, while it decays exponentially  in case of the social force model \cite{chraibi2015jamming}.
Velocity and distance-based interaction models also include extensions of the social force model with swarming mechanisms \cite{chen2018social,helbing2000simulating,lakoba2005modifications}. 
More recent models rely on the time-gap and the time-to-collision 
\cite{cordes2021time,karamouzas2014universal,moussaid2011simple}, temporal anticipation mechanisms \cite{lu2020pedestrian}, or anisotropic model with preferred crossing direction \cite{totzeck2020anisotropic}. 
For instance, the interaction of the universal power law model \cite{karamouzas2014universal} 
is based on a \textit{time-to-collision} variable calculated by linearly extrapolation of the trajectories of the pedestrians $i$ and $j$ based on their current velocities. Specifically, a collision is said to occur at some time $\tau_{ij} > 0$ if the discs modelling the pedestrian body intersect. 
If no collision occurs, the interaction force is zero. Of note is the fact that
this universal mechanism includes an interaction force and an anisotropic factor similar to the other models discussed above. 
This model is discussed in detail in Section~\ref{subsec:AntTTC}.
Even though the models are formulated as ordinary differential equations, the  forces proposed in the standard model do not have the regularity necessary for well-posedness in the classical sense of the theory of ordinary differential equations. Therefore, strong regularity assumptions need to be imposed to study the well-posedness of the systems mathematically, see for example \cite{canizo2011well,hauray2007n}.

As force-based models are second-order differential equations, they include inertia effects and relaxation. In the next section, we focus on the second major stream of continuous modelling of pedestrian dynamics: velocity-based models. These models are first-order approaches inspired from the robotics. They waive delay caused by inertia or relaxation.

\section{Velocity-based models}\label{sec:speed}

Although force-based models can describe many collective phenomena, they also have drawbacks. These are mainly due to the inertia which is inherent to second-order models. 
Examples of unrealistic behaviors are pedestrian overlaps, local oscillations, or tunneling effects, which occur especially in high density situations  \cite{chraibi2011force,sticco2020effects}. 
Additional specific hard-core forces with high expulsion rates allow to partly compensate for these undesired effects  \cite{helbing1995social,sticco2020effects}. 
Other approaches aim to model more realistic pedestrian shapes \cite{chraibi2011force,chraibi2010generalized}.
These extensions include additional parameters that can be difficult to calibrate.
Further drawbacks of force-based models pertain to numerical issues. 
Indeed, being of second order, force-based models can require additional computational complexity \cite{koster2013avoiding}. 
The assumption of force mechanisms governing pedestrian dynamics is also questionable \cite{chraibi2011force,tordeux2016collision}. 
The velocity of a pedestrian can be promptly adjusted at any time. 
Furthermore, step effects of the pedestrian motion showing oscillations make acceleration measurements imprecise.

Velocity-based pedestrian models are first-order approaches borrowed from robotics. 
A large class of velocity-based models relies on velocity optimization processes under collision constraints over given anticipation times. 
In contrast to force-based models that are by definition reactive, velocity-based models allow each agent to actively choose its next velocity using local anticipation mechanisms \cite{van2021algorithms}.
Most velocity-based models appeared later in the literature, from the end of the 2000s, partly with the objective of compensating for the shortcomings of the force-based models. 
Indeed, velocity-based approaches are rigid body models devoid of inertia for which it is straightforward to control hard-core exclusion, even for minimal models with few parameters.
Moreover, the simulation of first-order models requires less computational effort. 
We focus on the velocity obstacle model class in the following, before presenting a mathematical framework for collision-free dynamics and further hybrid approaches.

\subsection{Velocity obstacle models}
\label{sub:VO_models}
Velocity-based models are systems of first-order differential equations. 
They are given by
\begin{align}
    v_i=V_i(t,x,v),&&v_i(0)=v_{0,i},\quad i=1,\ldots,n,
\end{align}
with the velocity function $V_i\in\mathbb R^d(t,x,v)$ that can depend on the agent velocities $v$, making the model implicitly defined.
In certain models, the velocity function
\begin{align}
    V_i(t,x,v)=\omega_i(t,x,v)\,e_i(t,x,v),&&i=1,\ldots,n,
\end{align}
consists of a scalar speed model $\omega_i(t,x,v)\in\mathbb R$ and a pedestrian direction model $e_i(t,x,v)\in\mathbb R^d$.

Pioneer works in the field of velocity-based models date back to the end of the 1990s and the work of Fiorini et al.\ with velocity obstacle and velocity avoidance sets \cite{fiorini1998motion}. 
The approach consists of linearly extrapolating the trajectories of the pedestrians to determine collision sets. 
These sets describe the so-called  \textit{collision cones}. 
The pedestrian model is obtained by minimizing the deviation from the desired velocity outside the collision cones 
\begin{align}
    v_i=\text{arg}\hspace{-4mm}\min_{v\,\not\in\, \cup_{j\ne i}\text{VO}_{ij}}\|v-u_i\|^2,&&i=1,\ldots,n,
    \label{eq:VO1}
\end{align}
with $u_i$ the desired velocity and $\text{VO}_{ij}$ the collision cone of the $i$-th pedestrian with the $j$-th neighboring pedestrian. 
Here, the collision cones depend on the current velocities of the agents. 
This relationship makes the velocity model implicit. 
In practice, the implicit system is solved numerically using semi-implicit numerical schemes combining an explicit Euler solver for the velocity 
\begin{align*}
   v_i(t+dt)=\text{arg}\hspace{-5.5mm}\min_{v\,\not\in\, \cup_{j\ne i}\text{VO}_{ij}(t)}\|v-u_i(t)\|^2,&&i=1,\ldots,n,
\end{align*}
with an implicit Euler scheme for the positions of the agents \cite{berg2011reciprocal,van2008reciprocal}.
Note that in this case, the velocity model is equivalent to an explicit Euler discretisation of the second order model
\begin{align*}
    \dot v_i=\frac1\tau\big(\text{arg}\hspace{-4mm}\min_{v\,\not\in\, \cup_{j\ne i}\text{VO}_{ij}}\|v-u_i\|^2-v_i\big),&&i=1,\ldots,n,
\end{align*}
for a small inertial time $\tau$ equal to the discretisation time.

The velocity obstacle model may correspond to an evacuation behavior in which the pedestrians go as fast as possible to the destination.
The initial applications of the model of Fiorini et al.\ concern the movement of robots in an environment including moving obstacles and automated vehicles driving on a highway \cite{fiorini1998motion}. 
One of the first applications of velocity obstacle models for pedestrian dynamics, including model calibration and validation, is the work of Paris et al. \cite{paris2007pedestrian}. 
This study demonstrated the pertinence of VO models for pedestrian behaviors in case of evacuation. 
However, the VO models are of first order and they are implicitly defined with regard to the velocities. 
Their simulation can describe unrealistic oscillation effects for which two pedestrians oscillate in opposite directions (ping-pong). 
Extensions were proposed to overcome this undesired behavior.   
For instance, the \textit{reciprocal velocity obstacle} model (RVO) avoids oscillation by taking into account that neighbors use similar collision avoidance strategies \cite{van2008reciprocal}.
This formulation guarantees hard-core body exclusion under specific conditions.
The \textit{optimal reciprocal collision avoidance} model (ORCA) extends the collision-free dynamics to a general framework of agents acting independently without communicating with each other \cite{berg2011reciprocal}.
For more details on the ORCA model and alternatives we refer to  \cite{luo2018porca,narang2015generating,pellegrini2009you} and the reviews \cite{chraibi2018modelling,maury2018crowds,van2021algorithms}.

\subsection{General collision-free mathematical framework}

Parallel to the development of VO, RVO, and ORCA models, Venelle and Maury proposed a rigorous mathematical modelling framework for velocity-based approaches \cite{maury2008mathematical,maury2011discrete}. 
The set of admissible configurations of the system exclude overlapping configurations as it is defined as
\[
Q=\big\{x\in\mathbb R^{2N},~ D_{ij}=|x_j-x_i|-2r\ge0~\forall j\ne i\}.
\]
The set of admissible velocities constrains the system to remain in the admissible configurations as it is defined as
\[
C_x=\big\{v\in\mathbb R^{2N},~\forall j<i~ D_{ij}=0\Rightarrow \nabla D_{ij}\cdot v\ge0\big\}.
\]
If a contact occurs, that is, $D_{ij}=0$, the set of admissible velocities enforces motion away from the obstacle as $\nabla D_{ij}\cdot v\ge0$ implies that the distance increases. 
Indeed, $dD_{ij}/dt\ge0$ as $D_{ij}=0$ and, by continuity of the solutions, $D_{ij}$ remains positive. 
Moreover, if $x_0\in Q$ then $x$ remains in $Q$ for any $v \in C_x$. 
The actual velocity field is defined as the feasible field which is closest to the desired velocity field $u(x)$ in the least-square sense. This can be written as
\begin{equation}
\frac{dx}{dt}= P_{C_x}u(x)
\end{equation}
where $P_{C_x}$ denotes the Euclidean projection onto the closed convex cone $C_x$. 
Using the outward normal cone to the set of feasible configurations $Q$ allows us to reformulate the problem as a first-order differential inclusion \cite{maury2008mathematical}. 
As $Q$ is closed and convex, standard theory directly ensures well-posedness. 
In the general case where $Q$ is not convex, the uniform prox-regular property is suitable to provide well-posedness \cite{maury2008mathematical,maury2011discrete}. 
Most existing continuous first-order models belong to the velocity constraint proposed by Maury and Venel, including the velocity obstacle model class.
For instance, the scalar speed in the collision-free model \cite{tordeux2016collision} is a function $V$ of the smallest distance in front 
$D_i=\min_j D_{ij}$, 
with $j$ such that $\nabla D_{ij}\cdot v\le0$. 
Assume the velocity is zero in the obstacle direction in case of contact, i.e., $V(0)=0$, then the admissibility constraint $\nabla D_{ij}\cdot v\ge0$ holds and the model is collision-free.

The collision-free model may be referred to as a hybrid model, since the term modelling the direction is a superposition of repulsive pairwise gradient potentials  \cite{tordeux2016collision,xu2019generalized} like in the case of force-based models.
The \textit{gradient navigation} model has a similar direction term, while the speed model is a relaxation process of second order \cite{dietrich2014gradient}. 
In the synthetic vision model \cite{ondvrej2010synthetic}, the pedestrian direction is regulated according to the bearing angle and its time derivative. 
Indeed, the bearing angle remains constant as a collision occurs. 
The time-to-interaction is similar to the time-to-collision of force-based models and enables to regulate the speed. 
Generally speaking, the state-of-the-art velocity models aim to combine reactive physical mechanisms of force-based models with cognitive processes including anticipation and optimisation in order to develop general models providing accurate behaviors with parameters independent of the situations, e.g., different types of geometry or pedestrian density levels; more details on how this combination is achieved technically can be found in Sec.~\ref{sub:articulation}. 

\section{Anticipation in pedestrian models}\label{sec:anticipation}

\subsection{Different degrees of anticipation and planning}
The focus on the mechanism of collision avoidance in the models exposed in the previous sections
reflects the importance of having individual pedestrians adjust their velocities in the light of
how they anticipate that their neighbours will move. From a mathematical standpoint, the way in which
anticipation is handled strongly impacts the structure of the resulting model equations, along
an axis more or less orthogonal to the choice of first-order (velocity-based) or second-order (force-based) dynamics
for the adjustment of the velocity. To avoid duplicating the discussion for each of these two possibilities,
we will make use of the generalised cost of Van Toll et al. \cite{van2020generalized} to write the next velocity as the optimum of a generalised cost,
\begin{equation}
\centering
\displaystyle
v_i(t+dt)= \underset{v \in \mathbb{R}^2}{\mathrm{arg\, min}}\ C_i^{(t)}(v).
\label{eq:C_t}
\end{equation}
This is naturally compatible with the form of \eqref{eq:VO1}, provided that the cost $C_i^{(t)}$ is suitably modified to make velocity obstacles prohibitive.
In contrast, it should be noted that in other models \cite{karamouzas2017implicit,maury2008mathematical,maury2011discrete}, the minimisation is performed globally via a centralised algorithm, i.e., over a velocity set $\boldsymbol{v}=(v_1,\dots,v_N)\in \mathbb{R}^{2N}$ extending to all agents. In that case, as $v$ minimises a cost function $C^{(t)}$ over a multi-dimensional space involving all agents, it is akin to a system optimum, whereas the optimisation of Eq.~\ref{eq:C_t} is performed autonomously by each agent.

Regarding force-based models, it is worth recalling that the inertial term on the left-hand side of \eqref{eq:HT} can be recovered by differentiation, if $C_i^{(t)}$ includes an inertial contribution $m_i[v-v_i(t)]^2/2$ \cite{karamouzas2017implicit,van2020generalized}. No such contribution is present for velocity-based models.

These observations clarify that anticipation does not imply being able to foresee the future, i.e.,
requiring the introduction of future positions $r(t''>t)$ of other agents in the cost function $C_i^{(t)}$. Instead, agents
will try to predict these future positions on the basis of what they have observed at times $t'\leq t$.
When no such anticipation is at play, notably if $C_i^{(t)}$ only involves the current positions of other agents, the model is said to be reactive; this is notably the case of the simplified Social Force Model used in \cite{helbing2000simulating}.

\subsection{Anticipation based on velocity obstacles}
To go beyond strict reactiveness, the current velocities are introduced in $C_i^{(t)}$, whereby future positions can be extrapolated linearly with the help of $r(t'')=r(t)+(t''-t) v(t)$. 
In the VO models discussed in Sec.~\ref{sub:VO_models}, this extrapolation may be limited to a predefined, finite time horizon $T$, typically one or a few seconds into the future, $t'' \in [t, t+T]$ and any test velocity leading to an
(anticipated) collision over this time horizon, i.e., belonging to the truncated VO cone, is strictly prohibited.
Access is thus denied to the corresponding portion of phase space, so that, structurally, the problem turns into a geometric one, in which one must project the velocity as close as possible to the desired one via \eqref{eq:VO1}, while staying out of forbidden areas.
However, even when reciprocity issues (avoiding double avoidance moves) are properly handled, this finite horizon entails discontinuities and sharp turns, hence a somewhat robotic motion in some cases \cite{curtis2013pedestrian}.

\subsection{Anticipation based on times to collision}
\label{subsec:AntTTC}
Should one wish to extend the horizon indefinitely, say, at least until the next collision, then 
one should not bar collision-prone test velocities so strictly; otherwise, the whole test space $\mathbb{R}^2 \setminus \{0\}$ could become inaccessible. Instead,
it is sensible to penalize, rather than prohibit, velocities leading to  a collision, by defining a cost $E[\tau (v) ]$ based
on the anticipated time to collision $\tau(v)$ if velocity $v$ is selected. Interestingly, if agents are assumed to have a circular shape
of radius $r_i$, $\tau(v)$ can be expressed  explicitly 
as
\begin{equation}
     \tau_{ij}(v_i) =
     \begin{cases}
\frac{- \boldsymbol{x}_{ij}\cdot \boldsymbol{v}_{ij} - \sqrt{\Delta}}{v_{ij}^2} &\text{ if } \Delta \geqslant 0 \\
     \infty &\text{ otherwise},
\end{cases}
\end{equation}
where $\Delta=(x_{ij}\cdot v_{ij})^2 - v_{ij}^2[x_{ij}^2 - (\sigma_i + \sigma_j)^2]$, $x_{ij}$ and $v_{ij}$ are the relative positions and desired velocities of pedestrian $i$ with respect to pedestrian $j$.
Using  empirical pedestrian datasets, Karamouzas et al. \cite{karamouzas2014universal}  estimated that the cost should have the truncated power-law form 
\begin{equation} 
E(\tau)= A \frac{e^{-\tau/\tau_0}}{\tau^p} \text{ with }p\approx 2.
\end{equation}
Inserting the force that derives from
this energy into Newton's equation, a force-based model is thus obtained, which manages to effectively account for anticipation
over an unbounded time horizon while only using current positions and velocities.
Still, $E[\tau (v) ]$ exhibits strong discontinuities, insofar as the TTC jumps from a finite value for merely grazing trajectories to $\infty$ when the trajectory is deviated by the tiniest amount; grazing trajectories are then excessively favoured. This deficiency can be remedied by smoothing the transition from $\tau<\infty$  to $\tau=\infty$. This can be achieved by introducing some
uncertainty in the evaluation of the neighbour's radius or velocity \cite{karamouzas2017implicit}, or by considering that
encroachments into personal spaces should also be penalised \cite{echeverria2022anticipating}.

\subsection{Articulation between the cognitive layer and the mechanical layer}
\label{sub:articulation}
There remains the question of how to properly integrate the empirical \textit{anticipation} potential $E(\tau)$ into a force-based equation
that may also include mechanical forces. Traditionally, force-based models tended to turn a blind eye to the ontological difference
between forces and simply sum social forces and mechanical ones. On the other hand, VO models would overlook the existence
of mechanical forces.
A theoretically better grounded solution \cite{echeverria2022anticipating,hoogendoorn2003simulation,nicolas2020dense,van2020generalized} consists in separating the choice of a desired velocity from the mechanical equation governing its implementation. Thus, the minimisation performed in \eqref{eq:C_t} yields
the desired velocity, which is then be inserted into a mechanical equation and leads to
\begin{equation}
    v_i^{\mathrm{des}}(t+dt)= \underset{v \in \mathbb{R}^2}{\mathrm{arg\, min}}\,C_i^{(t)}(v) 
     \ \ \text{ and }\ \ 
    m_i \frac{dv_i}{dt} = m_i \frac{v^{\mathrm{des}}-v_i}{\tau} + F^{\mathrm{mech}}_{\to i}.
    \label{eq:coupled_cm}
\end{equation}
The differential structure of this equation is more sophisticated than only Newton's equation,
as it can integrate two distinct relaxation processes or delays, a psychological one and a mechanical one.
Although this conceptual decomposition has been largely overlooked in recent agent-based models, it is actually rooted in Hoogendoorn and Bovy's realisation in the early 2000s \cite{hoogendoorn2003simulation}, in light of Hill's earlier work, that "the task of guiding this vehicle is similar to the task of guiding a car, and can hence be described in an analogous manner", an analogy that emerged again in \cite{echeverria2022anticipating}.
Importantly, in contrast to \cite{karamouzas2017implicit}, the minimisation in \eqref{eq:coupled_cm} is performed by each (autonomous) agent,
and not collectively. 
Incidentally, the effects of inter-individual variations between agents can thus be clarified. For sure, differences in the anatomy of pedestrians introduce heterogeneity in the system, by notably affecting the mass $m_i$ and mechanical relaxation time $\tau$ entering the second half of \eqref{eq:coupled_cm}, as well as the contact forces $F^{\mathrm{mech}}_{\to i}$ (connected with agents' shapes). But heterogeneity naturally also affects the agents' preferences and behaviours, modelled in the cost function $C_i^{(t)}$; preferred speeds and avoidance behaviours are thus mostly controlled by this decisional layer, rather than by mechanics, even though they may be influenced by the anatomy, too.

Importantly, this formulation discards non-verbal cooperative mechanisms such as mutual gazes whereby agents
can negotiate how they will avoid each other. Recent experiments have however shown that 
mutual gazes were not critical for the natural avoidance of collisions between counter-walking pedestrians \cite{murakami2022spontaneous}.

\subsection{Differential games}
\label{sub:diff_games}
To further enhance the anticipation capabilities of the agents, the time horizon considered in the cost function $C_i^{(t)}$ of \eqref{eq:coupled_cm} can be extended, 
thereby becoming
a functional of the agent's test velocity $v_i(t'),\,t'\in [t,T_f]$ and the other agents' velocities 
$v_j(t'),\,t'\in [t,T_f]$. Schematically, the cost function then turns into \cite{bonnemain2022pedestrians,hoogendoorn2003simulation}

\begin{equation}
\centering
\displaystyle
C_i[ \mathcal{V} ]  = 
   \underset{\text{running cost}}
   {\underbrace{\int_{t}^{T_f}e_i\left[t', \mathcal{R}(t'),\mathcal{V}(t')\right] dt'}}
   +
   \underset{\text{terminal cost}}
   {\underbrace{E_i^{\mathrm{T}}\left[\mathcal{R}(T_f)\right]}},
 \label{eq:E_T_game}
\end{equation}
where $\mathcal{V}=(v_1, \dots ,v_N)$ denotes the velocities and $\mathcal{R}=(r_1, \dots ,r_N)$, the positions obtained by integrating $\mathcal{V}$ over time. (A discount factor can also be introduced into the integrand to have a stronger focus on the near future than on the long-term one.)
In performing this modification, we have altered the structure of the problem, in that
an agent will now act not only on the basis of observations, but also anticipation, that means, on this agent's guess of where the others will move, assuming that they reason like him or her. 
This marks the well-known, but nonetheless dramatic change from a mostly reactive mathematical structure to the game structure where players turn their reasoning towards the future. It was first explored by von Neumann and Morgenstern \cite{von2007theory} and leads e.g.
to Nash equilibria instead of minima of the global energy, with the additional difficulty
that the game is here formulated in a differential formalism.
In \eqref{eq:E_T_game}, the issue of inter-individual variability plays a role again and is made conspicuous by the subscripts $i$ in the cost functional $C_i$, the running cost $e_i$, and the terminal cost $E_i^{\mathrm{T}}$. Of course, such heterogeneity can be discarded, except perhaps in the terminal cost to account for the variety of destinations, to avoid the need to specify innumerable parameters. However, for evolutionary systems, it is known that even small heterogenities can
give rise to very substantial variations in the global outcome of the dynamics \cite{marsan2016stochastic}. Alternatively, they can be modelled as (frozen) noise in the cost functional $C_i$; the minimisation must then be performed with respect to expected (mean) values; this is at the heart of stochastic differential games \cite{marsan2016stochastic}, which can also account for the impact of exogeneous (time-dependent) noise on the dynamics.

Unfortunately, even when the functional forms of $e$ and $E_T$ are specified and, for instance, identical for all agents, the problem
is extremely complicated to solve and the computational cost associated therewith grows
exponentially with the number $N$ of \textit{players}. 
To circumvent this issue, one can settle with an exclusive focus on fairly simple scenarios \cite{hoogendoorn2003simulation} 
(also see \cite{zanardi2021urban} for recent examples with autonomous vehicles), 
or to evade the differential problem by discretising it and solving it with a cellular automaton, for instance in the context of egresses and evacuations 
\cite{bouzat2014game,heliovaara2013patient,von2015spatial}. Recently, confirming von Neumann and Morgenstern's very early intuition, an alternative has emerged in which the rest of the agents is handled in a
mean-field way, i.e., by considering their density rather than their individual positions \cite{bonnemain2022pedestrians,djehiche2017mean,jiang2016macroscopic,lachapelle2011mean,nasser2019crowd}. This typically leads to a system of two coupled equations, a backward mechanical one akin to \eqref{eq:coupled_cm} and a forward-looking one for the utility, which yields the control signal (i.e., the desired velocity) 
and may be recast into an Eikonal equation. It has very recently been shown that, unlike most existing agent-based models, these forward-backward equations can capture 
singular experimental features, notably observed during the crossing of a static crowd by an intruder \cite{bonnemain2022pedestrians}.



\section{Data-based calibration}\label{sec:calibration}
The models discussed in the previous sections are mainly mathematical or algorithmic representations of observations made by researchers while analyzing pedestrian interactions and studying pedestrian trajectories. Based on their perception, conclusions are drawn which are then modelled with the help of physical laws or algorithmic sequences driven by optimization problems. It is crucial to note that this procedure naturally filters information, as the human perception is limited. In contrast, data-based approaches build on artificial neural networks allow for modelling without any perception information or physical relationships. Of course, the training data chosen contains prior information and may induce a bias, which is the reason why the training data has to be chosen carefully. However, data-based approaches using artificial neural networks are an promising alternative modelling technique. 

Instead of going to the extremes, purely physical versus purely data-based modelling, we believe that combinations exploiting the advantages of each of the methods are reasonable. In fact, there are many shades of gray here. A first step to implement data in physical models is to fit the model parameters that are classically guessed from observations using optimization methods like for example in \cite{matei2019inferring,turarov2022gradient} on the microscopic level and \cite{gomes2019parameter} in a macroscopic setting. One step further is to replace certain terms of force-based models with neural networks. This was first investigated in \cite{gottlich2022parameter} for pedestrian and traffic dynamics and further studied in \cite{gaskin2022neural} for other social dynamics. The advantage of replacing only parts of the dynamics with neural networks is that the outcome of the models is comparable, see \cite{totzeck2022parameter}. If purely data-driven modelling is employed to predict whole trajectories, different information is used, which makes it difficult to find fair comparison measures.

In the following, we start from physically inspired models, where only the parameters are calibrated to data, then we proceed with the discussion of a hybrid model, where single force terms are replaced by feed-forward neural networks. An overview of pedestrian models that are only based on neural networks is given in section \ref{sec:preditive}.

\subsection{Data-based calibration of the model parameters}

Let us consider $N\in \mathbb N$ pedestrians with position and velocity trajectories \[x_i, v_i \colon [0,T] \rightarrow \mathbb R^d, \] respectively. A general force-based model is given by
\begin{subequations}\label{eq:state}
\begin{align}
\frac{d}{dt} x_i &= v_i, \qquad i=1,\dots,N, && x_i(0) = x_{0,i}, \\
\frac{d}{dt} v_i &= D_i(t,x_i,v_i) + I_i(t,x,v) + B(t,x_i,v_i), && v_i(0) = v_{0,i},
\end{align}
\end{subequations}
where $D_i \colon [0,T] \times \mathbb R^d \times \mathbb R^d \rightarrow \mathbb R^d $ encodes the desired velocity or destination of the $i$-th pedestrian, $I_i(t,x,v)$ denotes the interaction force acting on the $i$-th pedestrian and $B(t,x_i,v_i)$ allows to incorporate forces resulting from interactions with the surroundings, for example walls or columns in buildings. 

To give a flavor, we mention some simple examples for the three force terms: \[D_i(t,x_i,v_i) = \tau (u_i - v_i ), \] where $u_i$ is the desired velocity of pedestrian $i$ and $\tau>0$ a parameter that allows to adjust the velocity of relaxation towards this desired velocity; 
\[I_i(t,x,v) = -\frac{1}{N}\sum_{j=1}^N \nabla U_1(x_j - x_i),\] where $U_1 \in \mathcal C^1$ is a radially symmetric interaction potential which models binary interactions of the pedestrians; \[B(t,x_i,v_i) = -\frac{1}{M}\sum_{k=1}^M \nabla U_2(y_k - x_i),\] where $y_k$ is a discrete representation of walls and $U_2 \in \mathcal C^1$ is similar to $U_2$ with different parameters and describes the interaction of the $i$-th pedestrian with the wall. A feasible interaction potential is for example the Morse potential \cite{dorsogna2006self} \[U(d) = A e^{-|d|/a} - Re^{-|d|/r}, A,R \ge 0, a,r >0. \] However, for analytical results a smoothing is necessary to circumvent the singularity at zero. The parameters $A$ and $R$ allow to adjust the strength of attraction and repulsion while the ranges of the attraction and repulsion forces can be tuned by $a$ and $r$, respectively. We remark that the factors $1/N$ and $1/M$ can be neglected for purely microscopic models, but play an important role in the derivation of the mean-field equation.

Already the simple example above includes nine parameters, i.e.~$\tau, A_j, R_j, a_j, r_j$ for $j=1,2$. If the positions and possibly also the velocities of all (or a subset) of the pedestrians are available, techniques from optimal control can be employed for calibration. To this end, the formulation of an appropriate cost functional is necessary. In \cite{turarov2022gradient} a least square approach is chosen
\begin{equation*}
J(x,v,w) = \int_0^T \frac{\lambda_1}{2} | x(t) - x_\mathrm{data}(t) |^2 + \frac{\lambda_2}{2} | v(t) - v_\mathrm{data}(t) |^2 dt + \frac{\lambda_3}{2} |w -w_\mathrm{ref}|^2,
\end{equation*}
where $w =(\tau, A_1, R_1, a_1, r_1,A_2, R_2, a_2, r_2)$ is the parameter set and $w_\mathrm{ref}$ denotes reference parameters. If not applicable, the terms can be neglected by adjusting the weight parameters $\lambda_j\ge 0, j=1,2,3$. Of course there are many other choices for the cost functional. If the parameter set is constrained, we define the set of admissible parameters or controls, for example
\[
W_\mathrm{ad} = \{ w \in \mathbb R^9 \colon A_1,R_1,A_2,R_2 \ge 0,\quad  0 \le  \tau \}.
\]

Altogether, this leads to the optimization problem
\begin{equation}\label{opt}\tag{OP}
\min\limits_{w \in W_\mathrm{ad}} J(x,v,w) \quad \text{ subject to} \quad \eqref{eq:state}.
\end{equation}
Well-known techniques from optimal control with differential equations \cite{hinze2008optimization,troltzsch2010optimal} can be used to analyze the problem and derive calibration algorithms for numerical implementation, see for example \cite{gottlich2022parameter,turarov2022gradient}. Given sufficient regularity of the operators, the optimization can be structured as \textit{first optimize then discretize} based on adjoints. If this is infeasible, \textit{first discretize then optimize} methods for example with automatic differentiation (AD) can be employed to compute the gradient. If the parameter space is low dimensional, as in the example above, optimization techniques requiring only evaluations of $J$ are applicable as well, see \cite{totzeck2022parameter} for an example.

Note that the data are used here only to fit the parameters of the physically-inspired force terms. We emphasize that social interactions can only be approximated by physical forces. This can for example be illustrated with the help of binary interactions, see \cite{totzeck2020anisotropic}. There, the interaction of two pedestrians walking in opposite directions is considered. As the physical forces are aligned with the vector connecting the positions of the two pedestrians, they cannot escape from their desired path to avoid a collision and finally come to rest. 

Thus, the properties of the forces which are essential for physical applications limit the ability of the model to describe social interaction. In the next section, we therefore consider hybrid models, which replace the physically-inspired interaction terms with artificial neural networks and allow for richer interaction behaviours. 

\subsection{Data-based calibration of hybrid models}
Let us now consider hybrid models consisting of an ODE system to describe the evolution of the positions and velocities of each pedestrian with interaction forces resulting from neural networks. The idea behind this model is that data may contain more information about social interactions that can be perceived by humans and therefore have no chance to be considered in physics-based models.

To allow for a fair comparison of the hybrid model with the physics-based model of the previous subsection, we only replace parts of the force terms with neural networks. Indeed, we consider
\begin{subequations}\label{eq:hybrid}
\begin{align}
\frac{d}{dt} x_i &= v_i, \qquad i=1,\dots,N, && x_i(0) = x_{0,i}, \\
\frac{d}{dt} v_i &= D_i(t,x_i,v_i) + N_i(t,x,v), && v_i(0) = v_{0,i},
\end{align}
\end{subequations}
where $D_i(t,x_i,v_i)$ coincides with above and $N_i(t,x,v)$ is a neural network that models the interaction of the $i$-th pedestrian with his/her neighbors. One should underline that we learn one neural network that models the interaction with other pedestrians and walls at the same time. In fact, we cannot expect that neural networks split their behaviour intentionally into pedestrian interactions and wall interactions, which is also reported in  \cite{gottlich2022parameter}. We want to remark, that in contrast to so-called \textit{physics-informed neural networks}, the ODE structure is preserved by construction in the proposed dynamics. 

We emphasize that the general formulation with $N_i(t,x,v)$ allows to model binary interaction terms which are often used in crowd dynamics by setting
\[
N_i(t,x,v) = \sum_{j=1}^N g(t,x_i,v_i,x_j,v_j)
\]
and learn the function $g$. However, the framework above is more general and in particular the architecture of the neural networks is not fixed. In fact, it would be very interesting to study the influence of the neural network architecture on the outcome of the calibration. A suitable cost functional for the calibration is the one introduced above. 

A proof of concept of the hybrid approach was done in \cite{gottlich2022parameter, totzeck2022parameter} with different learning strategies. The first article uses a stochastic gradient descent approach, while the proceedings article works with Consensus-based optimization to calibrate the data. Both publications use small-scale neural networks, which are implemented by hand. For large-scale systems we recommend to use libraries such as \texttt{pytorch}.
The training data is taken from an experiment in Germany involving a crowd of pedestrians. Impressively, the forces predicted by the neural network for binary interactions of two pedestrians are far more realistic than the ones predicted by the calibrated physical model. In fact, the numerical study of binary interactions shows that the forces predicted by the neural net act orthogonally to the desired velocity, leading to natural collision avoidance behaviour. This is in contrast to the drawback of physics-inspired models discussed at the end of the last subsection. 

\medskip

To conclude this section, we want to remark that the data-based calibration influenced only the right-hand side of the ODE model which is used to predict the trajectories of positions and velocities of the pedestrians. Due to these structural assumptions, many results for ODE systems apply to the models discussed here. The models considered in the following are less restrictive. On the one hand, this allows us to incorporate more features hidden in the data. On the other hand, we lose mathematical structure which may be advantageous for rigorous analysis.

\section{Data-based predictive algorithms}\label{sec:preditive}
In the previous section, we discussed the hybrid approach, which combines physics-based and data-based approaches. In the following, we present a purely data-driven approach that utilizes neural networks, disregarding physical laws and prior knowledge.

Over the past decade, purely data-based approaches have gained significant attention for trajectory prediction tasks due to their real-world applications in areas such as autonomous vehicles and social robots \cite{korbmacher2022review}. These neural network based algorithms are particularly well-suited for these tasks, as they have demonstrated superior prediction accuracy when compared to physics-based models. One reason for that is their specific training to minimize Euclidean distance metrics such as average displacement error (ADE) and final displacement error (FDE) \cite{alahi2016social}, while physics-based models are primarily designed for other objectives. Another reason is that these neural networks possess many more parameters, which allows for more flexibility and a greater ability to learn a wider range of behaviors. A limitation of physics-based models is the pairwise modelling of interactions, we recall the isotropic interaction term of general force-based models  introduced in \eqref{eq:FBgeneral} as
\[
I_i(t,x,v) = -\sum_{j=1}^N \nabla U(x_j - x_i).
\]
The force modelling the pedestrian behavior results from the superposition of forces generated by pairwise interactions between the $i$-th pedestrian and successively the neighbours.

In the data-based approach, the pairwise modelling assumption of the interactions is not necessary. 
The predicted behavior is given by a general delay differential equation given by
\[v_i =\frac{d}{dt} x_i = N_i((x(u),v(u)),  u \in \mathbb [t-T_p,t]),
\]
where $T_p\ge0$ is a finite past horizon time.
The function $N_i$ often consists of neural networks including a large number of parameters, commonly referred to as weights. These weights allow the neural network to learn complex behaviors and interactions. The input to the neural network includes not only the current position $x$ and velocity $v$ of other pedestrians around the primary pedestrian $x_i$, but also the past positions $x(u)$ and velocities $v(u)$ over a discretized time interval. The neural network takes into account not just the current state, but also the past behavior, enabling more accurate predictions. This makes the differential equation delayed.
In the literature, the most common neural network architectures used for trajectory predictions are the long short-term memory (LSTM) networks and generative adversarial networks (GAN). 
These network families, training algorithms and applications to pedestrian trajectory prediction are presented in the following subsections.

\subsection{Long short-term memory networks}

Recurrent neural networks (RNNs) are machine learning algorithms particularly useful for tasks involving sequential data. This is because they use feedback connections to store a representation of recent input events as activations, which gives them a type of temporal memory \cite{li2018independently}. One well-known example of RNN is the LSTM network \cite{hochreiter1997long}, which uses special memory cells to store information. 
LSTMs perform successfully in tasks like speech recognition \cite{graves2013speech}, machine translation \cite{sutskever2014sequence}, handwriting recognition and generation, and image captioning \cite{vinyals2015show}. LSTM networks can be trained to generate sequences by processing real data one step at a time and predicting the next element in the sequence, ensuring temporal coherence. Although training LSTM networks can be time-consuming and challenging, they are efficient at processing new data in real time, making them a popular choice for tasks like tracking and prediction. 

Recently, LSTM networks have been applied to the task of predicting pedestrian trajectories. One notable example is the social-LSTM proposed by Alahi et al.\ \cite{alahi2016social}. 
The social-LSTM takes into account the interactions between pedestrians in crowded scenarios using a novel pooling layer called the social pooling layer. Other LSTM-based approaches for pedestrian trajectory prediction include models for additional scene information, such as the scene-LSTM \cite{manh2018scene}, and models that consider both the scene and the social interactions of pedestrians, like the social-scene LSTM \cite{xue2018ss}. Furthermore, some LSTM algorithms use attention mechanisms to determine the relative importance of each person in the scene \cite{fernando2018soft+}. Another approach is to use graph neural networks in combination with an LSTM algorithm. Each agent is treated as a node of a graph, which allows sharing information across the pedestrians as shown by Huang et al. \cite{huang2019stgat}.

\subsection{Generative adversarial networks}
A problem inherent with predictions of trajectories is
the multimodal nature of future pedestrian trajectories. Generative adversarial networks (GANs) have the potential to address this issue, as they can generate multiple potential outcomes. Using GANs, it is possible to predict a range of potential future trajectories, rather than just a single most likely trajectory \cite{salzmann2020trajectron++}. The architecture of a GAN includes a generator and a discriminator.  The generator creates different sample outputs from the training data while the discriminator attempts to distinguish between the generated samples and the true data distribution. The generator and discriminator are in a competition, similar to a two-player min-max game \cite{gupta2018social}. However, GANs can be challenging to train and may suffer from mode collapse \cite{arjovsky2017towards}. Applied for trajectory prediction, GANs are often combined with LSTM networks.

In the social GAN proposed by Gupta et al.\ \cite{gupta2018social}, the generator is composed of an encoder and decoder based on LSTM networks, as well as a context pooling module. The discriminator also utilizes LSTM networks. However, the pooling module in this algorithm assigns equal weight to all surrounding pedestrians, resulting in difficulty distinguishing the effects on the target pedestrian from pedestrians at different distances and speeds. To address this issue, many authors have incorporated attention mechanisms, such as the attention module added by Sadegehian et al.\ \cite{sadeghian2019sophie} to assign different weights to surrounding pedestrians and the static environment. Other algorithms, such as social ways \cite{amirian2019social} and social-BiGAT \cite{kosaraju2019social}, have also been proposed to improve the prediction of trajectories, especially in complex and diverse environments. Lv et al.\ \cite{lv2022improved} propose the use of transformer networks in combination with GANs to capture the uncertainty in predictions. To take the heterogeneity of road users into account, Lai et al.\ \cite{lai2020trajectory} propose an attention module with two components to address the complexity of scenes with multiple interacting agents of different types. 

Using neural networks for predicting pedestrian dynamics is a discipline that is rapidly evolving, with numerous innovative network architectures being introduced in recent years. However, from a microscopic perspective, we believe that these two particular architectures are the most noteworthy and relevant.



\section{Modelling development perspectives}\label{sec:perspectives}

In this final section, we highlight some time-continuous pedestrian modelling paradigms that we expect to thrive in the coming years.  
These development perspectives complement the different classes of models discussed throughout the chapter.

Recent investigations show that some swarm and force-based pedestrian models introduced in section~\ref{sec:force} can be formulated as port-Hamiltonian (pH) systems \cite{matei2019inferring,tordeux2022multi}. 
PH systems are modern modelling approaches of non-linear physical systems combining skew-symmetric and dissipative components with input and output ports. 
This decomposition of system dynamics is meaningful in many domains and is currently increasingly popular \cite{Rashad2020}. 
For pedestrian dynamics, the skew-symmetric component specific to the Hamiltonian structure can correspond to isotropic social forces. 
Information about  the pedestrians' desired velocity and sensitivity is provided via the input port, which can be interpreted as some kind of feedback control.
Besides identification of new fundamental physical modelling components, technical benefits of pH systems rely on 
direct Hamiltonian computation by energy balance. Non-linear stability analysis can employ the Hamiltonian as Lyapunov function \cite{jacob2023port}.
The pH framework opens the door for new modelling and analysis paradigms for pedestrian dynamics that we expect to grow in the future. 

Pedestrian models based on differential games currently seem to be a promising approach for collision avoidance.
In a differential game, pedestrians are no longer represented as particles subject to different types of forces but as players capable of developing strategies. 
Extending the model class detailed in section~\ref{sec:anticipation}, these strategies can assume anticipatory capabilities over indefinite periods and allow agent systems to reach specific stationary states, such as Nash equilibrium.
In a \textit{mean-field game}, rational agents for which individual costs depend on the others through a mean-field interaction lead in the mean-field limit to a backward Hamilton-–Jacobi–-Bellman equation coupled with a forward Fokker-–Planck equation \cite{huang2006large,lasry2007mean}. 
Pedestrian models by mean-field game were introduced in the early 2010s \cite{dogbe2010modeling,lachapelle2011mean} and are still an active research field 
likely to further expand in the future.

Furthermore, we see potential in the \emph{hybrid approach} described in section \ref{sec:calibration}. Although this approach is relatively new in the field, it holds the potential to overcome the limitations of both neural networks and physics-based models. Specifically, it aims to address the lack of explainability and generalization in neural networks, as well as the limited range different behaviors in physics-based models. In section \ref{sec:calibration}, data-based algorithms were used to improve models, but the other way around is also possible. This is often referred to as physics-informed neural networks (PINN) \cite{raissi2019physics}. One issue with the training of neural networks, is that data needs to be generated. One approach is to use  synthetic data obtained by simulation \cite{von2020combining}. Other interesting features are knowledge-guided design of the network architecture and loss function \cite{willard2022integrating}.

\paragraph{\bf Acknowledgments}  RK, AN and AT acknowledge the Franco-German research project MADRAS funded in France by the Agence Nationale de la Recherche (ANR, French National Research Agency), grant number ANR-20-CE92-0033, and in Germany by the Deutsche Forschungsgemeinschaft (DFG, German Research Foundation), grant number 446168800.

\bibliographystyle{acm} 
\bibliography{biblio}

\begin{thebibliography}{100}

\bibitem{alahi2016social}
{\sc Alahi, A., Goel, K., Ramanathan, V., Robicquet, A., Fei-Fei, L., and
  Savarese, S.}
\newblock Social lstm: Human trajectory prediction in crowded spaces.
\newblock In {\em Proceedings of the IEEE conference on computer vision and
  pattern recognition\/} (2016), pp.~961--971.

\bibitem{albi2019vehicular}
{\sc Albi, G., Bellomo, N., Fermo, L., Ha, S., Kim, J., Pareschi, L., Poyato,
  D., and Soler, J.}
\newblock Vehicular traffic, crowds, and swarms. on the kinetic theory approach
  towards research perspectives.
\newblock {\em Mathematical Models and Methods in Applied Sciences 29}, 10
  (2019), 1901--2005.

\bibitem{amirian2019social}
{\sc Amirian, J., Hayet, J.-B., and Pettr{\'e}, J.}
\newblock Social ways: Learning multi-modal distributions of pedestrian
  trajectories with gans.
\newblock In {\em Proceedings of the IEEE/CVF Conference on Computer Vision and
  Pattern Recognition Workshops\/} (2019), pp.~0--0.

\bibitem{arjovsky2017towards}
{\sc Arjovsky, M., and Bottou, L.}
\newblock Towards principled methods for training generative adversarial
  networks.
\newblock {\em arXiv preprint arXiv:1701.04862\/} (2017).

\bibitem{aylaj2020unified}
{\sc Aylaj, B., Bellomo, N., Gibelli, L., and Reali, A.}
\newblock A unified multiscale vision of behavioral crowds.
\newblock {\em Mathematical Models and Methods in Applied Sciences 30}, 01
  (2020), 1--22.

\bibitem{bailo2018pedestrian}
{\sc Bailo, R., Carrillo, J.~A., and Degond, P.}
\newblock Pedestrian models based on rational behaviour.
\newblock In {\em Crowd Dynamics, Volume 1}. Springer, 2018, pp.~259--292.

\bibitem{bellomo2022towards}
{\sc Bellomo, N., Gibelli, L., Quaini, A., and Reali, A.}
\newblock Towards a mathematical theory of behavioral human crowds.
\newblock {\em Mathematical Models and Methods in Applied Sciences 32}, 02
  (2022), 321--358.

\bibitem{bonnemain2022pedestrians}
{\sc Bonnemain, T., Butano, M., Bonnet, T., Echeverr{\'\i}a-Huarte, I., Seguin,
  A., Nicolas, A., Appert-Rolland, C., and Ullmo, D.}
\newblock Pedestrians in static crowds are not grains, but game players.
\newblock {\em Physical Review E 107}, 2 (2023), 024612.

\bibitem{bouzat2014game}
{\sc Bouzat, S., and Kuperman, M.~N.}
\newblock Game theory in models of pedestrian room evacuation.
\newblock {\em Physical Review E 89}, 3 (2014), 032806.

\bibitem{canizo2011well}
{\sc Canizo, J.~A., Carrillo, J.~A., and Rosado, J.}
\newblock A well-posedness theory in measures for some kinetic models of
  collective motion.
\newblock {\em Mathematical Models and Methods in Applied Sciences 21}, 03
  (2011), 515--539.

\bibitem{chen2018social}
{\sc Chen, X., Treiber, M., Kanagaraj, V., and Li, H.}
\newblock Social force models for pedestrian traffic--state of the art.
\newblock {\em Transport reviews 38}, 5 (2018), 625--653.

\bibitem{chraibi2015jamming}
{\sc Chraibi, M., Ezaki, T., Tordeux, A., Nishinari, K., Schadschneider, A.,
  and Seyfried, A.}
\newblock Jamming transitions in force-based models for pedestrian dynamics.
\newblock {\em Physical Review E 92}, 4 (2015), 042809.

\bibitem{chraibi2011force}
{\sc Chraibi, M., Kemloh, U., Schadschneider, A., and Seyfried, A.}
\newblock Force-based models of pedestrian dynamics.
\newblock {\em Networks \& Heterogeneous Media 6}, 3 (2011), 425.

\bibitem{chraibi2010generalized}
{\sc Chraibi, M., Seyfried, A., and Schadschneider, A.}
\newblock Generalized centrifugal-force model for pedestrian dynamics.
\newblock {\em Physical Review E 82}, 4 (2010), 046111.

\bibitem{chraibi2018modelling}
{\sc Chraibi, M., Tordeux, A., Schadschneider, A., and Seyfried, A.}
\newblock Modelling of pedestrian and evacuation dynamics.
\newblock {\em Encyclopedia of complexity and systems science\/} (2018), 1--22.

\bibitem{cordes2021time}
{\sc Cordes, J., Chraibi, M., Tordeux, A., and Schadschneider, A.}
\newblock Time-to-collision models for single-file pedestrian motion.
\newblock {\em Collective Dynamics 6\/} (2021), 1--10.

\bibitem{cristiani2014multiscale}
{\sc Cristiani, E., Piccoli, B., and Tosin, A.}
\newblock {\em Multiscale modeling of pedestrian dynamics}, vol.~12.
\newblock Springer, 2014.

\bibitem{curtis2013pedestrian}
{\sc Curtis, S.}
\newblock {\em Pedestrian velocity obstacles: Pedestrian simulation through
  reasoning in velocity space}.
\newblock PhD thesis, The University of North Carolina at Chapel Hill, 2013.

\bibitem{dietrich2014gradient}
{\sc Dietrich, F., and K{\"o}ster, G.}
\newblock Gradient navigation model for pedestrian dynamics.
\newblock {\em Physical Review E 89}, 6 (2014), 062801.

\bibitem{djehiche2017mean}
{\sc Djehiche, B., Tcheukam, A., and Tembine, H.}
\newblock A mean-field game of evacuation in multilevel building.
\newblock {\em IEEE Transactions on Automatic Control 62}, 10 (2017),
  5154--5169.

\bibitem{dogbe2010modeling}
{\sc Dogb{\'e}, C.}
\newblock Modeling crowd dynamics by the mean-field limit approach.
\newblock {\em Mathematical and Computer Modelling 52}, 9-10 (2010),
  1506--1520.

\bibitem{dorsogna2006self}
{\sc D'Orsogna, M.~R., Chuang, Y.-L., Bertozzi, A.~L., and Chayes, L.~S.}
\newblock Self-propelled particles with soft-core interactions: patterns,
  stability, and collapse.
\newblock {\em Physical review letters 96}, 10 (2006), 104302.

\bibitem{echeverria2022anticipating}
{\sc Echeverr{\'\i}a-Huarte, I., and Nicolas, A.}
\newblock Body and mind: Decoding the dynamics of pedestrians and the effect of
  smartphone distraction by coupling mechanical and decisional processes.
\newblock {\em Transportation Research Part C: Emerging Technologies 157\/}
  (2023), 104365.

\bibitem{fernando2018soft+}
{\sc Fernando, T., Denman, S., Sridharan, S., and Fookes, C.}
\newblock Soft+ hardwired attention: An lstm framework for human trajectory
  prediction and abnormal event detection.
\newblock {\em Neural networks 108\/} (2018), 466--478.

\bibitem{fiorini1998motion}
{\sc Fiorini, P., and Shiller, Z.}
\newblock Motion planning in dynamic environments using velocity obstacles.
\newblock {\em The international journal of robotics research 17}, 7 (1998),
  760--772.

\bibitem{gaskin2022neural}
{\sc Gaskin, T., Pavliotis, G.~A., and Girolami, M.}
\newblock Neural parameter calibration for large-scale multiagent models.
\newblock {\em Proceedings of the National Academy of Sciences 120}, 7 (2023),
  e2216415120.

\bibitem{gomes2019parameter}
{\sc Gomes, S.~N., Stuart, A.~M., and Wolfram, M.-T.}
\newblock Parameter estimation for macroscopic pedestrian dynamics models from
  microscopic data.
\newblock {\em SIAM Journal on Applied Mathematics 79}, 4 (2019), 1475--1500.

\bibitem{gottlich2022parameter}
{\sc G{\"o}ttlich, S., and Totzeck, C.}
\newblock Parameter calibration with stochastic gradient descent for
  interacting particle systems driven by neural networks.
\newblock {\em Mathematics of Control, Signals, and Systems 34}, 1 (2022),
  185--214.

\bibitem{graves2013speech}
{\sc Graves, A., Mohamed, A.-r., and Hinton, G.}
\newblock Speech recognition with deep recurrent neural networks.
\newblock In {\em 2013 IEEE international conference on acoustics, speech and
  signal processing\/} (2013), Ieee, pp.~6645--6649.

\bibitem{gupta2018social}
{\sc Gupta, A., Johnson, J., Fei-Fei, L., Savarese, S., and Alahi, A.}
\newblock Social gan: Socially acceptable trajectories with generative
  adversarial networks.
\newblock In {\em Proceedings of the IEEE conference on computer vision and
  pattern recognition\/} (2018), pp.~2255--2264.

\bibitem{hauray2007n}
{\sc Hauray, M., and Jabin, P.-E.}
\newblock N-particles approximation of the vlasov equations with singular
  potential.
\newblock {\em Archive for rational mechanics and analysis 183}, 3 (2007),
  489--524.

\bibitem{helbing1992fluid}
{\sc Helbing, D.}
\newblock A fluid-dynamic model for the movement of pedestrians.
\newblock {\em Complex Systems 6\/} (1992), 391--415.

\bibitem{helbing2005self}
{\sc Helbing, D., Buzna, L., Johansson, A., and Werner, T.}
\newblock Self-organized pedestrian crowd dynamics: Experiments, simulations,
  and design solutions.
\newblock {\em Transportation science 39}, 1 (2005), 1--24.

\bibitem{helbing2000simulating}
{\sc Helbing, D., Farkas, I., and Vicsek, T.}
\newblock Simulating dynamical features of escape panic.
\newblock {\em Nature 407}, 6803 (2000), 487--490.

\bibitem{helbing1995social}
{\sc Helbing, D., and Molnar, P.}
\newblock Social force model for pedestrian dynamics.
\newblock {\em Physical review E 51}, 5 (1995), 4282.

\bibitem{heliovaara2013patient}
{\sc Heli{\"o}vaara, S., Ehtamo, H., Helbing, D., and Korhonen, T.}
\newblock Patient and impatient pedestrians in a spatial game for egress
  congestion.
\newblock {\em Physical Review E 87}, 1 (2013), 012802.

\bibitem{henderson1971statistics}
{\sc Henderson, L.}
\newblock The statistics of crowd fluids.
\newblock {\em Nature 229\/} (1971), 381--383.

\bibitem{hinze2008optimization}
{\sc Hinze, M., Pinnau, R., Ulbrich, M., and Ulbrich, S.}
\newblock {\em Optimization with PDE constraints}, vol.~23.
\newblock Springer Science \& Business Media, 2008.

\bibitem{hirai1975simulation}
{\sc Hirai, K., and Tarui, K.}
\newblock A simulation of the behavior of a crowd in panic.
\newblock In {\em Proceedings of the 1975 International Conference on
  Cybernetics and Society\/} (1975), pp.~409--411.

\bibitem{hochreiter1997long}
{\sc Hochreiter, S., and Schmidhuber, J.}
\newblock Long short-term memory.
\newblock {\em Neural computation 9}, 8 (1997), 1735--1780.

\bibitem{hoogendoorn2003simulation}
{\sc Hoogendoorn, S., and HL~Bovy, P.}
\newblock Simulation of pedestrian flows by optimal control and differential
  games.
\newblock {\em Optimal control applications and methods 24}, 3 (2003),
  153--172.

\bibitem{huang2006large}
{\sc Huang, M., Malham{\'e}, R.~P., and Caines, P.~E.}
\newblock Large population stochastic dynamic games: closed-loop mckean-vlasov
  systems and the nash certainty equivalence principle.
\newblock {\em Communications in Information \& Systems 6}, 3 (2006), 221--252.

\bibitem{huang2019stgat}
{\sc Huang, Y., Bi, H., Li, Z., Mao, T., and Wang, Z.}
\newblock Stgat: Modeling spatial-temporal interactions for human trajectory
  prediction.
\newblock In {\em Proceedings of the IEEE/CVF International Conference on
  Computer Vision\/} (2019), pp.~6272--6281.

\bibitem{jacob2023port}
{\sc Jacob, B., and Totzeck, C.}
\newblock Port-hamiltonian structure of interacting particle systems and its
  mean-field limit.
\newblock {\em arXiv:2301.06121\/} (2023).

\bibitem{jiang2016macroscopic}
{\sc Jiang, Y.-Q., Guo, R.-Y., Tian, F.-B., and Zhou, S.-G.}
\newblock Macroscopic modeling of pedestrian flow based on a second-order
  predictive dynamic model.
\newblock {\em Applied Mathematical Modelling 40}, 23-24 (2016), 9806--9820.

\bibitem{karamouzas2014universal}
{\sc Karamouzas, I., Skinner, B., and Guy, S.~J.}
\newblock Universal power law governing pedestrian interactions.
\newblock {\em Physical review letters 113}, 23 (2014), 238701.

\bibitem{karamouzas2017implicit}
{\sc Karamouzas, I., Sohre, N., Narain, R., and Guy, S.~J.}
\newblock Implicit crowds: Optimization integrator for robust crowd simulation.
\newblock {\em ACM Transactions on Graphics (TOG) 36}, 4 (2017), 1--13.

\bibitem{korbmacher2022review}
{\sc Korbmacher, R., and Tordeux, A.}
\newblock Review of pedestrian trajectory prediction methods: Comparing deep
  learning and knowledge-based approaches.
\newblock {\em IEEE Transactions on Intelligent Transportation Systems\/}
  (2022).

\bibitem{kosaraju2019social}
{\sc Kosaraju, V., Sadeghian, A., Mart{\'\i}n-Mart{\'\i}n, R., Reid, I.,
  Rezatofighi, H., and Savarese, S.}
\newblock Social-bigat: Multimodal trajectory forecasting using bicycle-gan and
  graph attention networks.
\newblock {\em Advances in Neural Information Processing Systems 32\/} (2019).

\bibitem{koster2013avoiding}
{\sc K{\"o}ster, G., Treml, F., and G{\"o}del, M.}
\newblock Avoiding numerical pitfalls in social force models.
\newblock {\em Physical Review E 87}, 6 (2013), 063305.

\bibitem{lachapelle2011mean}
{\sc Lachapelle, A., and Wolfram, M.-T.}
\newblock On a mean field game approach modeling congestion and aversion in
  pedestrian crowds.
\newblock {\em Transportation research part B: methodological 45}, 10 (2011),
  1572--1589.

\bibitem{lai2020trajectory}
{\sc Lai, W.-C., Xia, Z.-X., Lin, H.-S., Hsu, L.-F., Shuai, H.-H., Jhuo, I.-H.,
  and Cheng, W.-H.}
\newblock Trajectory prediction in heterogeneous environment via attended
  ecology embedding.
\newblock In {\em Proceedings of the 28th ACM International Conference on
  Multimedia\/} (2020), pp.~202--210.

\bibitem{lakoba2005modifications}
{\sc Lakoba, T.~I., Kaup, D.~J., and Finkelstein, N.~M.}
\newblock Modifications of the helbing-molnar-farkas-vicsek social force model
  for pedestrian evolution.
\newblock {\em Simulation 81}, 5 (2005), 339--352.

\bibitem{lasry2007mean}
{\sc Lasry, J.-M., and Lions, P.-L.}
\newblock Mean field games.
\newblock {\em Japanese journal of mathematics 2}, 1 (2007), 229--260.

\bibitem{li2010stability}
{\sc Li, H., Zhao, X., and Xie, D.}
\newblock Stability analysis of pedestrian flow and phase structure in the
  improved two-dimensional ov models.
\newblock In {\em 2010 Third International Joint Conference on Computational
  Science and Optimization\/} (2010), vol.~1, IEEE, pp.~161--165.

\bibitem{li2018independently}
{\sc Li, S., Li, W., Cook, C., Zhu, C., and Gao, Y.}
\newblock Independently recurrent neural network (indrnn): Building a longer
  and deeper rnn.
\newblock In {\em Proceedings of the IEEE conference on computer vision and
  pattern recognition\/} (2018), pp.~5457--5466.

\bibitem{lu2020pedestrian}
{\sc L{\"u}, Y.-X., Wu, Z.-X., and Guan, J.-Y.}
\newblock Pedestrian dynamics with mechanisms of anticipation and attraction.
\newblock {\em Physical Review Research 2}, 4 (2020), 043250.

\bibitem{luo2018porca}
{\sc Luo, Y., Cai, P., Bera, A., Hsu, D., Lee, W.~S., and Manocha, D.}
\newblock Porca: Modeling and planning for autonomous driving among many
  pedestrians.
\newblock {\em IEEE Robotics and Automation Letters 3}, 4 (2018), 3418--3425.

\bibitem{lv2022improved}
{\sc Lv, Z., Huang, X., and Cao, W.}
\newblock An improved gan with transformers for pedestrian trajectory
  prediction models.
\newblock {\em International Journal of Intelligent Systems 37}, 8 (2022),
  4417--4436.

\bibitem{manh2018scene}
{\sc Manh, H., and Alaghband, G.}
\newblock Scene-lstm: A model for human trajectory prediction.
\newblock {\em arXiv preprint arXiv:1808.04018\/} (2018).

\bibitem{marsan2016stochastic}
{\sc Marsan, G.~A., Bellomo, N., and Gibelli, L.}
\newblock Stochastic evolutionary differential games toward a systems theory of
  behavioral social dynamics.
\newblock {\em Mathematical Models and Methods in Applied Sciences 26}, 06
  (2016), 1051--1093.

\bibitem{martinez2017modeling}
{\sc Martinez-Gil, F., Lozano, M., Garc{\'\i}a-Fern{\'a}ndez, I., and
  Fern{\'a}ndez, F.}
\newblock Modeling, evaluation, and scale on artificial pedestrians: a
  literature review.
\newblock {\em ACM Computing Surveys (CSUR) 50}, 5 (2017), 1--35.

\bibitem{matei2019inferring}
{\sc Matei, I., Mavridis, C., Baras, J.~S., and Zhenirovskyy, M.}
\newblock Inferring particle interaction physical models and their dynamical
  properties.
\newblock In {\em 2019 IEEE 58th Conference on Decision and Control (CDC)\/}
  (2019), IEEE, pp.~4615--4621.

\bibitem{maury2018crowds}
{\sc Maury, B., and Faure, S.}
\newblock {\em Crowds in Equations: An Introduction to the Microscopic Modeling
  of Crowds}.
\newblock World Scientific, 2018.

\bibitem{maury2008mathematical}
{\sc Maury, B., and Venel, J.}
\newblock A mathematical framework for a crowd motion model.
\newblock {\em Comptes Rendus Mathematique 346}, 23-24 (2008), 1245--1250.

\bibitem{maury2011discrete}
{\sc Maury, B., and Venel, J.}
\newblock A discrete contact model for crowd motion.
\newblock {\em ESAIM: Mathematical Modelling and Numerical Analysis 45}, 1
  (2011), 145--168.

\bibitem{moussaid2011simple}
{\sc Moussa{\"\i}d, M., Helbing, D., and Theraulaz, G.}
\newblock How simple rules determine pedestrian behavior and crowd disasters.
\newblock {\em Proceedings of the National Academy of Sciences 108}, 17 (2011),
  6884--6888.

\bibitem{murakami2022spontaneous}
{\sc Murakami, H., Tomaru, T., Feliciani, C., and Nishiyama, Y.}
\newblock Spontaneous behavioral coordination between avoiding pedestrians
  requires mutual anticipation rather than mutual gaze.
\newblock {\em Iscience 25}, 11 (2022), 105474.

\bibitem{nakayama2005instability}
{\sc Nakayama, A., Hasebe, K., and Sugiyama, Y.}
\newblock Instability of pedestrian flow and phase structure in a
  two-dimensional optimal velocity model.
\newblock {\em Physical Review E 71}, 3 (2005), 036121.

\bibitem{nakayama2008effect}
{\sc Nakayama, A., Hasebe, K., and Sugiyama, Y.}
\newblock Effect of attractive interaction on instability of pedestrian flow in
  a two-dimensional optimal velocity model.
\newblock {\em Physical Review E 77}, 1 (2008), 016105.

\bibitem{narang2015generating}
{\sc Narang, S., Best, A., Curtis, S., and Manocha, D.}
\newblock Generating pedestrian trajectories consistent with the fundamental
  diagram based on physiological and psychological factors.
\newblock {\em PLoS one 10}, 4 (2015), e0117856.

\bibitem{nasser2019crowd}
{\sc Nasser, N., El~Ouadrhiri, A., El~Kamili, M., Ali, A., and Anan, M.}
\newblock Crowd management services in hajj: a mean-field game theory approach.
\newblock In {\em 2019 IEEE Wireless Communications and Networking Conference
  (WCNC)\/} (2019), IEEE, pp.~1--7.

\bibitem{nicolas2020dense}
{\sc Nicolas, A.}
\newblock Dense pedestrian crowds versus granular packings: An analogy of
  sorts.
\newblock In {\em Traffic and Granular Flow 2019\/} (2020), Springer,
  pp.~411--419.

\bibitem{ondvrej2010synthetic}
{\sc Ond{\v{r}}ej, J., Pettr{\'e}, J., Olivier, A.-H., and Donikian, S.}
\newblock A synthetic-vision based steering approach for crowd simulation.
\newblock {\em ACM Transactions on Graphics (TOG) 29}, 4 (2010), 1--9.

\bibitem{paris2007pedestrian}
{\sc Paris, S., Pettr{\'e}, J., and Donikian, S.}
\newblock Pedestrian reactive navigation for crowd simulation: a predictive
  approach.
\newblock In {\em Computer Graphics Forum\/} (2007), vol.~26(3), Wiley Online
  Library, pp.~665--674.

\bibitem{pellegrini2009you}
{\sc Pellegrini, S., Ess, A., Schindler, K., and Van~Gool, L.}
\newblock You'll never walk alone: Modeling social behavior for multi-target
  tracking.
\newblock In {\em 2009 IEEE 12th international conference on computer vision\/}
  (2009), IEEE, pp.~261--268.

\bibitem{raissi2019physics}
{\sc Raissi, M., Perdikaris, P., and Karniadakis, G.~E.}
\newblock Physics-informed neural networks: A deep learning framework for
  solving forward and inverse problems involving nonlinear partial differential
  equations.
\newblock {\em Journal of Computational physics 378\/} (2019), 686--707.

\bibitem{Rashad2020}
{\sc Rashad, R., Califano, F., van~der Schaft, A.~J., and Stramigioli, S.}
\newblock Twenty years of distributed port-hamiltonian systems: a literature
  review.
\newblock {\em IMA Journal of Mathematical Control and Information 37}, 4
  (2020), 1400--1422.

\bibitem{sadeghian2019sophie}
{\sc Sadeghian, A., Kosaraju, V., Sadeghian, A., Hirose, N., Rezatofighi, H.,
  and Savarese, S.}
\newblock Sophie: An attentive gan for predicting paths compliant to social and
  physical constraints.
\newblock In {\em Proceedings of the IEEE/CVF conference on computer vision and
  pattern recognition\/} (2019), pp.~1349--1358.

\bibitem{salzmann2020trajectron++}
{\sc Salzmann, T., Ivanovic, B., Chakravarty, P., and Pavone, M.}
\newblock Trajectron++: Dynamically-feasible trajectory forecasting with
  heterogeneous data.
\newblock In {\em European Conference on Computer Vision\/} (2020), Springer,
  pp.~683--700.

\bibitem{song2018simulation}
{\sc Song, X., Xie, H., Sun, J., Han, D., Cui, Y., and Chen, B.}
\newblock Simulation of pedestrian rotation dynamics near crowded exits.
\newblock {\em IEEE Transactions on Intelligent Transportation Systems 20}, 8
  (2018), 3142--3155.

\bibitem{sticco2020effects}
{\sc Sticco, I.~M., Frank, G.~A., and Dorso, C.~O.}
\newblock Effects of the body force on the pedestrian and the evacuation
  dynamics.
\newblock {\em Safety science 129\/} (2020), 104829.

\bibitem{sutskever2014sequence}
{\sc Sutskever, I., Vinyals, O., and Le, Q.~V.}
\newblock Sequence to sequence learning with neural networks.
\newblock {\em Advances in neural information processing systems 27\/} (2014).

\bibitem{tordeux2016collision}
{\sc Tordeux, A., Chraibi, M., and Seyfried, A.}
\newblock Collision-free speed model for pedestrian dynamics.
\newblock In {\em Traffic and Granular Flow'15}. Springer, 2016, pp.~225--232.

\bibitem{tordeux2022multi}
{\sc Tordeux, A., and Totzeck, C.}
\newblock Multi-scale description of pedestrian collective dynamics with
  port-hamiltonian systems.
\newblock {\em Networks and Heterogeneous Media 18}, 2 (2023), 906--929.

\bibitem{totzeck2020anisotropic}
{\sc Totzeck, C.}
\newblock An anisotropic interaction model with collision avoidance.
\newblock {\em Kinetic \& Related Models 13}, 6 (2020), 1219--1242.

\bibitem{totzeck2022parameter}
{\sc Totzeck, C., and G{\"o}ttlich, S.}
\newblock Parameter calibration with consensus-based optimization for
  interaction dynamics driven by neural networks.
\newblock In {\em European Consortium for Mathematics in Industry\/} (2022),
  Springer, pp.~17--22.

\bibitem{troltzsch2010optimal}
{\sc Tr{\"o}ltzsch, F.}
\newblock {\em Optimal control of partial differential equations: theory,
  methods, and applications}, vol.~112.
\newblock American Mathematical Soc., 2010.

\bibitem{turarov2022gradient}
{\sc Turarov, Z., and Totzeck, C.}
\newblock Gradient-based parameter calibration of an anisotropic interaction
  model for pedestrian dynamics.
\newblock {\em European Journal of Applied Mathematics\/} (2023), 1–22.

\bibitem{berg2011reciprocal}
{\sc van~den Berg, J., Guy, S.~J., Lin, M., and Manocha, D.}
\newblock Reciprocal n-body collision avoidance.
\newblock In {\em Robotics research}. Springer, 2011, pp.~3--19.

\bibitem{van2008reciprocal}
{\sc Van~den Berg, J., Lin, M., and Manocha, D.}
\newblock Reciprocal velocity obstacles for real-time multi-agent navigation.
\newblock In {\em 2008 IEEE international conference on robotics and
  automation\/} (2008), Ieee, pp.~1928--1935.

\bibitem{van2020generalized}
{\sc van Toll, W., Grzeskowiak, F., Gand{\'\i}a, A.~L., Amirian, J., Berton,
  F., Bruneau, J., Daniel, B.~C., Jovane, A., and Pettr{\'e}, J.}
\newblock Generalized microscropic crowd simulation using costs in velocity
  space.
\newblock In {\em Symposium on Interactive 3D Graphics and Games\/} (2020),
  pp.~1--9.

\bibitem{van2021algorithms}
{\sc Van~Toll, W., and Pettr{\'e}, J.}
\newblock Algorithms for microscopic crowd simulation: Advancements in the
  2010s.
\newblock In {\em Computer Graphics Forum\/} (2021), vol.~40(2), Wiley Online
  Library, pp.~731--754.

\bibitem{vermuyten2016review}
{\sc Vermuyten, H., Beli{\"e}n, J., De~Boeck, L., Reniers, G., and Wauters, T.}
\newblock A review of optimisation models for pedestrian evacuation and design
  problems.
\newblock {\em Safety science 87\/} (2016), 167--178.

\bibitem{vinyals2015show}
{\sc Vinyals, O., Toshev, A., Bengio, S., and Erhan, D.}
\newblock Show and tell: A neural image caption generator.
\newblock In {\em Proceedings of the IEEE conference on computer vision and
  pattern recognition\/} (2015), pp.~3156--3164.

\bibitem{von2007theory}
{\sc Von~Neumann, J., and Morgenstern, O.}
\newblock Theory of games and economic behavior.
\newblock In {\em Theory of games and economic behavior}. Princeton university
  press, 1944.

\bibitem{von2020combining}
{\sc von Rueden, L., Mayer, S., Sifa, R., Bauckhage, C., and Garcke, J.}
\newblock Combining machine learning and simulation to a hybrid modelling
  approach: Current and future directions.
\newblock In {\em International Symposium on Intelligent Data Analysis\/}
  (2020), Springer, pp.~548--560.

\bibitem{von2015spatial}
{\sc von Schantz, A., and Ehtamo, H.}
\newblock Spatial game in cellular automaton evacuation model.
\newblock {\em Physical Review E 92}, 5 (2015), 052805.

\bibitem{willard2022integrating}
{\sc Willard, J., Jia, X., Xu, S., Steinbach, M., and Kumar, V.}
\newblock Integrating scientific knowledge with machine learning for
  engineering and environmental systems.
\newblock {\em ACM Computing Surveys 55}, 4 (2022), 1--37.

\bibitem{xu2019generalized}
{\sc Xu, Q., Chraibi, M., Tordeux, A., and Zhang, J.}
\newblock Generalized collision-free velocity model for pedestrian dynamics.
\newblock {\em Physica A: Statistical Mechanics and its Applications 535\/}
  (2019), 122521.

\bibitem{xue2018ss}
{\sc Xue, H., Huynh, D.~Q., and Reynolds, M.}
\newblock Ss-lstm: A hierarchical lstm model for pedestrian trajectory
  prediction.
\newblock In {\em 2018 IEEE Winter Conference on Applications of Computer
  Vision (WACV)\/} (2018), IEEE, pp.~1186--1194.

\bibitem{yu2005centrifugal}
{\sc Yu, W., Chen, R., Dong, L.-Y., and Dai, S.}
\newblock Centrifugal force model for pedestrian dynamics.
\newblock {\em Physical Review E 72}, 2 (2005), 026112.

\bibitem{zanardi2021urban}
{\sc Zanardi, A., Mion, E., Bruschetta, M., Bolognani, S., Censi, A., and
  Frazzoli, E.}
\newblock Urban driving games with lexicographic preferences and socially
  efficient nash equilibria.
\newblock {\em IEEE Robotics and Automation Letters 6}, 3 (2021), 4978--4985.

\end{thebibliography}

\end{document}